\documentclass[aps,twocolumn,nofootinbib,superscriptaddress]{revtex4-2}

\usepackage[utf8]{inputenc}
\usepackage[T1]{fontenc}
\usepackage{amsmath,amssymb,amsfonts}
\usepackage{graphicx}
\usepackage{booktabs}
\usepackage{orcidlink}
\usepackage{url}
\hypersetup{colorlinks=true,linkcolor=blue,citecolor=blue,urlcolor=blue,
            pdftitle={Descartes Curvature Geometry, Compact-Cycle Amplitudes, and the GST Cabibbo Seed},
            pdfauthor={Andrew M. Brilliant}}

\begin{document}

\title{Descartes Curvature Geometry, Compact-Cycle Amplitudes, and the GST Cabibbo Seed}

\author{Andrew M. Brilliant \protect\orcidlink{0009-0004-8024-5442}}

\affiliation{Applied Dynamics Research, Sapporo, Japan}

\email{ab@ad-research.org}

\begin{abstract}
Several apparently distinct formulations of the charged-lepton mass relation
$p_2/e_1^2 = 2/3$ share a common algebraic object: the ratio of
flavor-breaking to flavor-symmetric amplitude components on an internal family
space. We show that (i)~the relation expressed as a Fourier-mode constraint on
Shulga's compact cycle, (ii)~the same relation as a Descartes curvature
condition in the sense of Kocik, and (iii)~the Soddy completion of the
resulting circle configuration are related by exact algebra, with no
approximations. The Soddy completion defines a companion scale
$\mathcal{F}^2 = \alpha^2\mu_\star$ with
$\alpha = \sqrt{3/2}-1$ and $\mu_\star = \sum_\ell m_\ell$; identifying
this scale with $m_s^{\overline{\text{MS}}}(\mu_\star)$ yields a light-quark mass
cascade with the ratio $m_d/m_s = 5/2 - \sqrt{6}$. Combined with the
leading Gatto--Sartori--Tonin relation, this gives the down-sector Cabibbo
seed $\sqrt{m_d/m_s} = \sqrt{3/2} - 1 = 0.22474\ldots$, numerically close to
the measured $|V_{us}| = 0.22503 \pm 0.00068$ ($0.4\sigma$). A full CKM
evaluation requires the up-sector phase correction, which we defer. The
correspondence is testable by future lattice QCD+QED determinations of
$m_d/m_s$; explicit failure modes are listed.
\end{abstract}

\maketitle


\section{Introduction}\label{sec:intro}

The Cabibbo angle has been measured to high precision through the CKM element
$|V_{us}|$, with the current world average\cite{PDG2026}
$|V_{us}| = 0.22503 \pm 0.00068$. The Gatto--Sartori--Tonin (GST)
relation\cite{GST1968,Fritzsch1979}
\begin{equation}\label{eq:GST}
  \sin\theta_C \;\approx\; \sqrt{\frac{m_d}{m_s}}\,,
\end{equation}
connects this angle to the ratio of down- and strange-quark masses. Gatto,
Sartori, and Tonin originally derived the Cabibbo angle from the cancellation
of quadratically divergent weak self-masses;\cite{GST1968} the quark-mass-ratio
form~(\ref{eq:GST}) is due to the Fritzsch texture-zero
ansatz.\cite{Fritzsch1979} Equation~(\ref{eq:GST}) is well established; what
remains unexplained, within this phenomenological relation, is why the ratio
$m_d/m_s$ has its observed value.

Separately, the empirical relation
\begin{equation}\label{eq:lepton_condition}
  \frac{\sum_\ell m_\ell}
  {\bigl(\sum_\ell \sqrt{m_\ell}\bigr)^2}
  \;=\; \frac{2}{3}
\end{equation}
has held for the charged-lepton pole masses to one part in $10^5$ since it was
first noted by Koide.\cite{Koide1982} Several authors have recast this relation
in different mathematical
languages,\cite{Kocik2012,Zenczykowski2012,Shulga2026} raising the
question of whether these are independent observations or coordinate
descriptions of a single underlying structure.

In this paper we pursue two aims. First, we show that the Fourier-mode
constraint of Shulga's compact-cycle model,\cite{Shulga2026} the Descartes
curvature reading of Kocik,\cite{Kocik2012} and the Soddy completion of the
resulting circle configuration are related by exact algebra, with no
approximations (Sec.~\ref{sec:equivalence}). Second, we show that identifying
the Soddy companion scale with the strange-quark mass fixes $m_d/m_s$ as a
geometric constant, and that the leading GST relation then yields a
closed-form down-sector Cabibbo seed (Sec.~6). 

\section{The Algebraic Object}\label{sec:algebraic}

Let $e_1 = \sum_i \sqrt{m_i}$, $p_2 = \sum_i m_i$ for the three charged
leptons, and define the lepton-sum scale
$\mu_\star \equiv p_2 = 1883.1$~MeV. The
condition~(\ref{eq:lepton_condition}) is
$p_2 = \frac{2}{3}e_1^2$, equivalently
$e_1 = \sqrt{3/2}\;\sqrt{p_2}$.

Define the companion amplitude
\begin{equation}\label{eq:F}
  \mathcal{F} \;\equiv\; e_1 - \sqrt{p_2}
  \;=\; \alpha\,\sqrt{\mu_\star}\,,
\end{equation}
where
\begin{equation}\label{eq:alpha}
  \alpha \;\equiv\; \sqrt{\tfrac{3}{2}} - 1 \;\approx\; 0.2247\,.
\end{equation}
The squared companion is
\begin{equation}\label{eq:F2}
  \mathcal{F}^2 \;=\; \alpha^2\,\mu_\star\,,
  \qquad \alpha^2 = \tfrac{5}{2} - \sqrt{6}
  \approx 0.0505\,.
\end{equation}
This is the central algebraic object. The three-input symmetric-polynomial
identity~(\ref{eq:lepton_condition}) collapses to one dimensionless constant
$\alpha$ times one mass scale $\mu_\star$, with $\alpha$ fixed by the
condition~(\ref{eq:lepton_condition}).

\section{Equivalence of Descriptions}\label{sec:equivalence}

We now show that three apparently distinct formulations of the
condition~(\ref{eq:lepton_condition}) describe the same constraint.

\subsection{Compact-cycle amplitudes (Shulga)}

Shulga\cite{Shulga2026} constructs a real amplitude on an internal circle,
\begin{equation}\label{eq:Z}
  Z(\phi) \;=\; c_0 + 2r\cos(\phi + \delta)\,,
\end{equation}
built from the two lowest antiperiodic modes, with the three charged-lepton
families sampling at equally spaced points:
$\sqrt{m_a} = Z(2\pi a/3)$ for $a = 0,1,2$. The parameter $c_0$ is the
flavor-symmetric (democratic) component and $r$ the flavor-breaking amplitude;
the cosine coefficient is $2r$, so Shulga's notation
$Z = s[1+\sqrt{2}\cos(\phi+\theta)]$ corresponds to $c_0 = s$ and
$2r/c_0 = \sqrt{2}$, i.e.\ $r/c_0 = 1/\sqrt{2}$.

By the orthogonality of equally spaced phases on the circle,
$\sum_a \cos(2\pi a/3 + \delta) = 0$ for any $\delta$, and similarly for the
double-angle sum. Therefore
\begin{align}
  e_1 &= \sum_a Z(2\pi a/3) = 3c_0\,, \label{eq:e1_shulga}\\
  p_2 &= \sum_a Z(2\pi a/3)^2 = 3(c_0^2 + 2r^2)\,. \label{eq:p2_shulga}
\end{align}
The ratio $p_2/e_1^2$ becomes
\begin{equation}\label{eq:Q_shulga}
  \frac{p_2}{e_1^2} = \frac{c_0^2 + 2r^2}{3c_0^2}
  = \frac{1}{3} + \frac{2r^2}{3c_0^2}\,.
\end{equation}
Setting this equal to $2/3$ gives
\begin{equation}\label{eq:shulga_constraint}
  \frac{r}{c_0} = \frac{1}{\sqrt{2}}\,.
\end{equation}
The condition~(\ref{eq:lepton_condition}) is exactly the statement that the
flavor-breaking and flavor-symmetric amplitudes on the compact cycle stand in
the ratio $1:\sqrt{2}$. Shulga derives this ratio from the spinor-squaring of
antiperiodic modes and a reality condition on the
amplitude.\cite{Shulga2026}

\subsection{Descartes curvatures (Kocik)}

The Descartes circle theorem\cite{Coxeter1968} for four mutually tangent
circles reads $(\sum_i k_i)^2 = 2\sum_i k_i^2$.
Kocik\cite{Kocik2012} observed that identifying curvatures
$k_i = \sqrt{m_i}$ places~(\ref{eq:lepton_condition}) within the generalized
Descartes family. The present work uses the standard (mutually tangent)
theorem and computes the fourth curvature.

The outer Soddy curvature of the three-circle configuration is
$k_4^- = e_1 - 2\sqrt{e_2}$, where
$e_2 = \sum_{i<j}\sqrt{m_i m_j}$.
Under~(\ref{eq:lepton_condition}), one has $4e_2 = p_2$, so
$k_4^- = e_1 - \sqrt{p_2} = \mathcal{F}$: the
Soddy curvature \textit{is} the companion amplitude~(\ref{eq:F}).

\subsection{The equivalence}

Expressing the Soddy curvature in Shulga's variables
via~(\ref{eq:e1_shulga})--(\ref{eq:shulga_constraint}):
\begin{equation}\label{eq:F_shulga}
  \mathcal{F} = e_1 - \sqrt{p_2}
  = 3c_0 - \sqrt{6}\,c_0 = c_0(3 - \sqrt{6})\,.
\end{equation}
Then $\alpha = \mathcal{F}/\sqrt{\mu_\star}
= c_0(3-\sqrt{6})/\sqrt{6c_0^2} = (3-\sqrt{6})/\sqrt{6}
= \sqrt{3/2} - 1$, confirming~(\ref{eq:alpha}).

The three descriptions are related by exact algebra:
\begin{center}
\tabcolsep10pt
\begin{tabular}{@{}lll@{}}
\toprule
Framework & Constraint & Language \\
\midrule
Shulga & $r/c_0 = 1/\sqrt{2}$ & Fourier modes \\[2pt]
Condition~(\ref{eq:lepton_condition})
  & $p_2/e_1^2 = 2/3$ & Sym.\ polynomials \\[2pt]
Soddy & $\mathcal{F} = \alpha\sqrt{\mu_\star}$
  & Inversive geometry \\
\bottomrule
\end{tabular}
\end{center}
No approximations are involved in any direction. The constant
$\alpha = \sqrt{3/2}-1$ appears in each language as the unique dimensionless
number implied by the shared constraint. The plus branch
$e_1 + \sqrt{p_2}$ gives $\sqrt{3/2}+1 \approx 2.22$, which has no
interpretation in the quark sector; the minus branch is the only
phenomenologically relevant root.

\section{The Mass Cascade}\label{sec:cascade}

The squared companion $\mathcal{F}^2 = \alpha^2\mu_\star$ evaluates to
$95.11$~MeV. Reference~\cite{Brilliant2026} established
this coincidence using PDG/FLAG~2024 inputs
($m_s^{\overline{\text{MS}}}(\mu_\star)
= 95.08 \pm 0.69$~MeV, $+0.06\sigma$); updated to
PDG~2026\cite{PDG2026} inputs and four-loop running,\cite{CRunDec}
$m_s^{\overline{\text{MS}}}(\mu_\star) = 94.5 \pm 0.7$~MeV, and the
coincidence holds at $+0.9\sigma$.
Identifying $\mathcal{F}^2$ with $m_s$ motivates the cascade
\begin{align}
  m_s&\stackrel{\text{ansatz}}{=} \alpha^2\mu_\star\,,\label{eq:ms}\\
  m_d&\stackrel{\text{ansatz}}{=} \alpha^4\mu_\star\,,\label{eq:md}\\
  m_u&\stackrel{\text{ansatz}}{=} \alpha^2\sqrt{2m_e\,\mu_\star}\,,\label{eq:mu}
\end{align}
where~(\ref{eq:ms}) is the Soddy identification,~(\ref{eq:md}) extends the
geometric progression by one step of $\alpha^2$,
and~(\ref{eq:mu}) sets
$m_u = \sqrt{m_d \cdot 2m_e}$ (the geometric mean of $m_d$ and the pair
threshold). A full derivation
of~(\ref{eq:mu}) from the Descartes framework requires further work.
The cascade reproduces PDG 2026 light-quark
$\overline{\text{MS}}$ masses at $\mu_\star$ within current
uncertainties;\cite{PDG2026,Brilliant2026} the strange-quark prediction is
within $1\sigma$ ($+0.9\sigma$),
while $m_u$ and $m_d$ are consistent within their larger fractional
uncertainties.

\section{The GST Cabibbo Seed}\label{sec:cabibbo}

From Eqs.~(\ref{eq:ms})--(\ref{eq:md}),
\begin{equation}\label{eq:ratio}
  \frac{m_d}{m_s} = \alpha^2 = \frac{5}{2} - \sqrt{6}\,.
\end{equation}
Substituting into the leading GST relation~(\ref{eq:GST}),
\begin{equation}\label{eq:main}
  \boxed{\;\sqrt{\frac{m_d}{m_s}} \;=\; \alpha \;=\;
    \sqrt{\tfrac{3}{2}} - 1 \;=\; 0.22474\ldots\;}
\end{equation}
This is the down-sector Cabibbo seed in the sense of the GST/Fritzsch
leading-order relation. Numerically it lies $0.4\sigma$ from the measured
$|V_{us}| = 0.22503 \pm 0.00068$.

The full GST expression including the up-sector
correction,\cite{Fritzsch1979}
\begin{equation}\label{eq:full_GST}
  |V_{us}| \;\approx\; \left|\sqrt{\frac{m_d}{m_s}} -
  e^{i\phi}\sqrt{\frac{m_u}{m_c}}\right|,
\end{equation}
includes a correction of order
$\sqrt{m_u/m_c} \approx 0.04$, of the right sign and magnitude to account for
the residual. A precise evaluation requires extending the cascade to the
charm sector, which we defer to future work.

\section{The Fritzsch Texture}\label{sec:fritzsch}

The GST relation arises from the four-zero texture\cite{Fritzsch1979}
\begin{equation}\label{eq:texture}
  M_d = \begin{pmatrix}
    0 & A & 0 \\ A^* & 0 & B \\ 0 & B^* & C
  \end{pmatrix}.
\end{equation}
In the cascade framework, the eigenvalues
of~(\ref{eq:texture}) are expressed through
$\alpha$ and $\mu_\star$; the expansion parameter is
$\alpha = \sqrt{3/2}-1$, selected by the Descartes geometry rather than
fitted.
The cascade interpolates between two physical scales:
$\mu_\star = 1883$~MeV
and $2m_e = 1.022$~MeV:
\begin{equation}\label{eq:ladder}
  \mu_\star \;\xrightarrow{\;\alpha^2\;}\;
  m_s \;\xrightarrow{\;\alpha^2\;}\; m_d\,,\qquad
  m_u = \sqrt{m_d \cdot 2m_e}\,.
\end{equation}
The step size of this geometric ladder is $\alpha^2$---the
square of the GST Cabibbo seed.

\section{Weak-Basis Considerations}\label{sec:weak}

The condition~(\ref{eq:lepton_condition}) holds for charged-lepton pole
masses, where the mass and weak bases are approximately aligned: in the
Standard Model, the charged-lepton Yukawa matrix can be taken diagonal without
loss of generality when neutrino masses are neglected. The amplitude structure
described in Sec.~\ref{sec:equivalence} is therefore visible directly in the
physical spectrum.

For quarks the situation is different. The CKM matrix rotates between up-type
and down-type mass bases, so the condition~(\ref{eq:lepton_condition}) should
not be expected to hold for quark pole masses directly. The
cascade~(\ref{eq:ms})--(\ref{eq:mu}) operates on
$\overline{\text{MS}}$ running masses at a specific scale, and the
GST relation~(\ref{eq:GST}) itself arises from the Fritzsch texture in the
weak basis. Whether the amplitude structure of
Sec.~\ref{sec:equivalence} is native to the weak-basis Yukawa matrix, or
requires a more involved embedding, is an open question.

In this reading, the charged-lepton identity~(\ref{eq:lepton_condition}) and
the leading GST factor are two projections of the same constant~$\alpha$,
expressed in the lepton and quark sectors respectively.

\section{Tests and Failure Modes}\label{sec:tests}

The correspondence can be falsified if:

\noindent (i) future lattice QCD+QED determinations move
$m_s^{\overline{\text{MS}}}(\mu_\star)$ outside the companion value
$\mathcal{F}^2 = 95.11 \pm 0.01$~MeV, or move $m_d/m_s$ away from
$\alpha^2 = 0.0505$;

\noindent (ii) a full weak-basis GST/Fritzsch treatment, including the
up-sector phase correction, fails to reproduce $|V_{us}|$ without additional
fitted structure;

\noindent (iii) Shulga's compact-cycle mechanism cannot be extended beyond
charged leptons in a way consistent with the cascade;

\noindent (iv) the weak-basis embedding of Sec.~\ref{sec:weak} proves
inconsistent with CKM data;

\noindent (v) or the apparent agreement disappears under consistent QCD+QED
scheme conversion and running between the lepton pole-mass and quark
$\overline{\text{MS}}$ sides of the comparison.

The lepton-side quantity $\mathcal{F}^2$ is fixed to better than $0.01\%$; each
future FLAG compilation\cite{FLAG2024} is therefore a direct test. A future
determination with $\sigma_{m_s} < 0.3$~MeV would provide a decisive test. The
present result should be viewed as a suggestive numerical correspondence whose
status will be determined by future data.

\section{Related Work}\label{sec:related}

Rivero\cite{Rivero2011} showed independently that chained algebraic conditions
connect lepton and quark sectors and recover $m_s$ in the same numerical
neighborhood via a different route.
\.Zenczykowski\cite{Zenczykowski2012} developed a $Z_3$-symmetric
parametrization of fermion mass square roots that
embeds the condition~(\ref{eq:lepton_condition}) in a broader framework
including quarks and neutrinos, with the phase $\delta \simeq 2/9$ playing a
central role. Both approaches are compatible with, but not derived from, the
equivalence established in Sec.~\ref{sec:equivalence}.

\section{Summary}\label{sec:summary}

The charged-lepton mass condition $p_2/e_1^2 = 2/3$, the Fourier-mode
constraint $r/c_0 = 1/\sqrt{2}$ on Shulga's compact cycle, and the
Descartes--Soddy curvature relation
$\mathcal{F} = \alpha\sqrt{\mu_\star}$ are the same algebraic constraint in
different coordinates, related by exact algebra. Identifying the Soddy
companion scale
with $m_s^{\overline{\text{MS}}}(\mu_\star)$ yields the ratio
$m_d/m_s = 5/2 - \sqrt{6}$,
and the leading Gatto--Sartori--Tonin relation gives the down-sector Cabibbo
seed $\sqrt{m_d/m_s} = \sqrt{3/2} - 1 = 0.22474$, within $0.4\sigma$ of the
measured $|V_{us}| = 0.22503$.

A single constant $\alpha = \sqrt{3/2}-1$ appears as the companion norm-deficit
associated with the compact-cycle amplitude constraint, the light-quark mass
hierarchy parameter, the Fritzsch texture expansion parameter, and the leading
GST Cabibbo seed. The convergence is empirical: the same constant appears in
four distinct phenomenological contexts. Whether this reflects a common origin
is an open question that future precision measurements will decide.

\section*{Acknowledgments}
The author thanks A.~Rivero for extensive correspondence on the algebraic
connections between lepton and quark sectors, the lattice QCD community for
successive improvements to light-quark mass determinations, and K.~Shulga for
the compact-cycle construction that clarifies the amplitude-level content of
the lepton mass relation.

\emph{AI usage disclosure.} AI tools were used for programming assistance and
document formatting only. All mathematical derivations, physical arguments,
numerical predictions, and scientific claims originate from the author or from
prior published work.

\end{document}